\documentclass[physrev,reprint,superscriptaddress,bibnotes]{revtex4-2}  

\usepackage[T1]{fontenc}
\usepackage[utf8]{inputenc} 
\usepackage[australian]{babel}

\usepackage[a4paper,centering,hmargin=1.75cm,vmargin=2cm]{geometry} 
\usepackage{amsmath,amssymb,graphicx,bm,microtype,xcolor} 
\usepackage[colorlinks,allcolors=blue!50!black]{hyperref} 
\usepackage[all]{hypcap}
\usepackage{cleveref,braket,siunitx,booktabs}
\usepackage[version=4]{mhchem}

\begin{document}

\title{Experimental Quantum Simulation of Chemical Dynamics}

\author{Tomas~Navickas}
\affiliation{School of Physics, University of Sydney, NSW 2006, Australia}
\affiliation{ARC Centre of Excellence for Engineered Quantum Systems, University of Sydney, NSW 2006, Australia}

\author{Ryan J.~MacDonell}
\affiliation{ARC Centre of Excellence for Engineered Quantum Systems, University of Sydney, NSW 2006, Australia}
\affiliation{School of Chemistry, University of Sydney, NSW 2006, Australia}
\affiliation{Departments of Chemistry and Physics, Dalhousie University, Halifax, NS B3H 4R2, Canada}

\author{Christophe H.~Valahu}
\affiliation{School of Physics, University of Sydney, NSW 2006, Australia}
\affiliation{ARC Centre of Excellence for Engineered Quantum Systems, University of Sydney, NSW 2006, Australia}
\affiliation{University of Sydney Nano Institute, University of Sydney, NSW 2006, Australia}

\author{Vanessa C.~Olaya-Agudelo}
\affiliation{ARC Centre of Excellence for Engineered Quantum Systems, University of Sydney, NSW 2006, Australia}
\affiliation{School of Chemistry, University of Sydney, NSW 2006, Australia}

\author{Frank~Scuccimarra}
\affiliation{School of Physics, University of Sydney, NSW 2006, Australia}
\affiliation{ARC Centre of Excellence for Engineered Quantum Systems, University of Sydney, NSW 2006, Australia}

\author{Maverick J.~Millican}
\affiliation{School of Physics, University of Sydney, NSW 2006, Australia}
\affiliation{ARC Centre of Excellence for Engineered Quantum Systems, University of Sydney, NSW 2006, Australia}

\author{Vassili G.~Matsos}
\affiliation{School of Physics, University of Sydney, NSW 2006, Australia}
\affiliation{ARC Centre of Excellence for Engineered Quantum Systems, University of Sydney, NSW 2006, Australia}

\author{Henry L.~Nourse}
\affiliation{School of Chemistry, University of Sydney, NSW 2006, Australia}

\author{Arjun~D.~Rao}
\affiliation{School of Physics, University of Sydney, NSW 2006, Australia}
\affiliation{ARC Centre of Excellence for Engineered Quantum Systems, University of Sydney, NSW 2006, Australia}

\author{Michael J.~Biercuk}
\affiliation{School of Physics, University of Sydney, NSW 2006, Australia}
\affiliation{ARC Centre of Excellence for Engineered Quantum Systems, University of Sydney, NSW 2006, Australia}

\author{Cornelius~Hempel}
\affiliation{School of Physics, University of Sydney, NSW 2006, Australia}
\affiliation{ARC Centre of Excellence for Engineered Quantum Systems, University of Sydney, NSW 2006, Australia}
\affiliation{ETH Zurich-PSI Quantum Computing Hub, Laboratory for Nano and Quantum Technologies (LNQ), Paul Scherrer Institut, 5232 Villigen, Switzerland}

\author{Ivan~Kassal}
\email{ivan.kassal@sydney.edu.au}
\affiliation{School of Chemistry, University of Sydney, NSW 2006, Australia}
\affiliation{ARC Centre of Excellence for Engineered Quantum Systems, University of Sydney, NSW 2006, Australia}
\affiliation{University of Sydney Nano Institute, University of Sydney, NSW 2006, Australia}

\author{Ting Rei~Tan}
\email{tingrei.tan@sydney.edu.au}
\affiliation{School of Physics, University of Sydney, NSW 2006, Australia}
\affiliation{ARC Centre of Excellence for Engineered Quantum Systems, University of Sydney, NSW 2006, Australia}
\affiliation{University of Sydney Nano Institute, University of Sydney, NSW 2006, Australia}

\begin{abstract}
\begin{center}
    \textbf{Abstract}
\end{center}

Accurate simulation of dynamical processes in molecules and reactions is among the most challenging problems in quantum chemistry. Quantum computers promise efficient chemical simulation, but the existing quantum algorithms require many logical qubits and gates, placing practical applications beyond existing technology. Here, we carry out the first quantum simulations of chemical dynamics by employing a more hardware-efficient encoding scheme that uses both qubits and bosonic degrees of freedom. Our trapped-ion device accurately simulates the dynamics of non-adiabatic chemical processes, which are among the most difficult problems in computational chemistry because they involve strong coupling between electronic and nuclear motions. We demonstrate the programmability and versatility of our approach by simulating the dynamics of three different molecules as well as open-system dynamics in the condensed phase, all with the same quantum resources. Our approach requires orders of magnitude fewer resources than equivalent qubit-only quantum simulations, demonstrating the potential of using hybrid encoding schemes to accelerate quantum simulations of complex chemical processes, which could have applications in fields ranging from energy conversion and storage to biology and drug design.

\end{abstract}

\maketitle

\section{Introduction}

Dynamics is central to chemistry because it underpins all of chemical reactivity and kinetics. However, predicting the quantum-mechanical dynamics of both nuclei and electrons involved in chemical transformations remains one of the great challenges of computational chemistry. Particularly difficult are simulations of non-adiabatic processes, where the breakdown of the Born-Oppenheimer approximation leads to the entangled motion of nuclear wavepackets on multiple potential-energy surfaces on ultrafast timescales~\cite{domcke04,cederbaum81}. These effects are especially important in photochemistry, including almost all chemical reactions in the atmosphere, such as those responsible for smog formation and ozone depletion; interactions of organisms with light, whether through photosynthesis, vision, or UV DNA damage; solar energy conversion through solar cells; and compounds for preventing and treating disease, from sunscreens to photodynamic therapies~\cite{Forbes2015}.

\begin{figure*}[t!]
    \centering
    \includegraphics[]{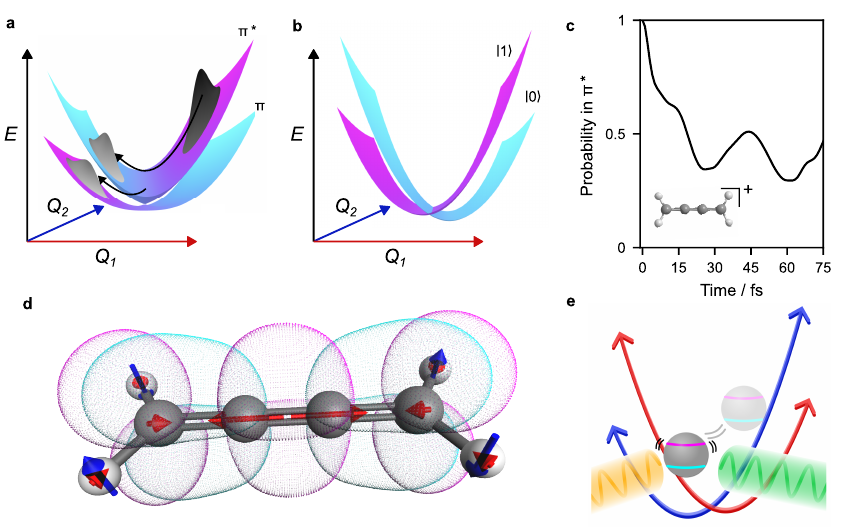}
    \caption{\textbf{Mapping non-adiabatic chemical dynamics onto an MQB simulator.} 
    \textbf{(a)}~Adiabatic potential energy surfaces~(PES) of the butatriene cation determined from solving the Born-Oppenheimer Hamiltonian at all nuclear positions. The cyan and magenta regions of the surfaces correspond to $\pi$ and $\pi$* electronic characters. The two PESs intersect at a conical intersection~(CI) through a coupling with the central bond stretching ($Q_1$) and torsion ($Q_2$) vibrational modes. This vibronic coupling allows ultrafast molecular dynamics: an initial wavepacket (grey, top right) on the upper surface approaches the conical intersection and splits into two entangled branches on the two electronic surfaces. 
    \textbf{(b)}~Diabatic representation of the PESs. The coupling between surfaces (not shown) is linear along $Q_2$.
    \textbf{(c)}~Probability of the wavepacket remaining in the initial diabatic state $\pi^*$ as a function of time.
    \textbf{(d)}~Butatriene molecule with its electronic states represented by orbitals (pink and cyan) that differ between the two states, and the two vibrational modes $Q_1$ (red) and $Q_2$ (dark blue). 
    \textbf{(e)}~Molecular electronic and vibrational degrees of freedom can both be mapped onto a mixed qudit-boson (MQB) simulator consisting here of one trapped ion. Potentials and vibronic couplings are induced with lasers (yellow and green beams).
    } 
    \label{fig:generaloverview}
\end{figure*}

Methods to simulate non-adiabatic dynamics on conventional, classical computers are either approximate or limited by computational scaling that is unpredictable a~priori~\cite{domcke04,worth08}. Even more complicated is simulating open quantum systems, including chemical dynamics in the condensed phase~\cite{Nitzan}. On classical computers, doing so usually involves tracking density matrices (more complicated than wavefunctions) as well as the system's interactions with its environment. Classical multi-configuration time-dependent Hartree (MCTDH) calculations have been achieved for relatively small systems with explicit bath modes~\cite{worth98,worth08,bonfanti12} or weak system-bath couplings~\cite{vanhaeften23}, but increasing the coupling strength or the number of modes quickly becomes intractable.

Simulating chemistry on quantum computers promises to overcome the challenges faced by conventional computational techniques. Indeed, using quantum machines for simulating nature is the idea that launched quantum computing~\cite{Feynman1982}, because a controllable quantum device could mimic another quantum system efficiently (with polynomial resources in system size)~\cite{Lloyd1996}. Simulations of quantum chemistry, encompassing both electronic structure and dynamics, are particularly suited for quantum simulation and are likely to be the earliest applications of quantum computing~\cite{Lanyon2010,Cao2019,McArdle2020}. Most chemical quantum algorithms have focused on finding static molecular properties, usually energies~\cite{Aspuru2005,Lanyon2010,Peruzzo2014,Kandala2017,Hempel2018,Cao2019,Bauer2020,Nam2020}. By contrast, there are few quantum algorithms for molecular dynamics~\cite{Kassal2008, Sawaya2019, Ollitrault2020,Miessen2023}, despite the centrality of dynamics to chemistry. To make matters worse, useful applications of those algorithms would require quantum computers with many more low-error qubits and gates than is feasible with existing technology~\cite{Kassal2008,Ollitrault2020,Miessen2023}.

The hardware costs of existing algorithms can be reduced by encoding information not only in qubits but in other degrees of freedom as well~\cite{MacDonell2021,crane2024}. This approach is particularly natural in the context of chemical dynamics, where molecular vibrational degrees of freedom can be mapped onto bosonic hardware elements present in multiple quantum architectures~\cite{MacDonell2021,MacDonell2023,Wang2023,Valahu2023,Whitlow2023,So2024,Kang2024,Sun2024,olayaagudelo2024arXiv}. Using bosons reduces hardware resource requirements compared to the alternative of using a large number of qubits to encode a single continuous degree of freedom. In particular, the mixed-qudit-boson (MQB) approach allows simulations of molecular models with vibrational-electronic (vibronic) couplings using significantly reduced quantum resources compared to qubit-only simulations~\cite{MacDonell2021,olayaagudelo2024arXiv}. This is achieved using qudits ($d$-level quantum systems) controllably coupled to bosonic modes to encode, respectively, molecular electronic states and nuclear vibrations (Fig.~\ref{fig:generaloverview}). MQB simulators based on trapped-ion systems have been used to predict vibronic spectra and to observe the dynamical geometric-phase interference around an engineered conical intersection~\cite{Shen2018, MacDonell2023,Valahu2023,Whitlow2023}. Previous experimental quantum simulations of chemical dynamics were important but limited demonstrations, restricted to non-molecular model Hamiltonians or being non-programmable. A programmable MQB simulator would allow tunable experimental parameters to simulate a wide range of molecules using the same hardware. 

Here, we perform the first programmable MQB simulations of non-adiabatic photochemical dynamics using a trapped-ion system. We demonstrate the programmability of the simulator by reproducing the real-time molecular dynamics of three different photoexcited molecules---the allene cation, the butatriene cation, and pyrazine---encoded using different parameters in a vibronic-coupling Hamiltonian. In each molecule, the vibronic couplings give rise to a conical intersection, which allows ultrafast (femtosecond) population transfer between electronic states~\cite{domcke04,cederbaum81}. Experimental measurements reliably reproduce the expected chemical dynamics, including signatures of the conical intersections. We further demonstrate the extensibility of our simulator to open-system dynamics by implementing simulations of pyrazine coupled to a thermal bath, showing the expected damping of coherent effects and thermalization. Our results further demonstrate the hardware efficiency of MQB simulation: with one trapped-ion qudit, two vibrational modes, and one laser pulse, it achieves a computation that would require 11 qubits and over $10^5$ entangling gates in a standard qubit-only framework.

\section{Methods}

\subsection{MQB Simulation of Photochemistry}

In an ion-trap MQB simulator, molecular vibrations and electronic states are encoded in motional and electronic degrees of freedom of the trapped ion~\cite{MacDonell2021} (Fig.~\ref{fig:generaloverview}). We simulate non-adiabatic dynamics in an analog fashion by reconstructing the time evolution of important molecular properties, rescaled to a timescale accessible by the simulator. Typically, temporal dynamics are rescaled from femtoseconds to milliseconds---i.e., by a factor of approximately $10^{11}$---sufficient to enable direct probing with conventional electronics and laser systems.

To represent molecular systems, MQB simulators use vibronic coupling (VC) Hamiltonians~\cite{MacDonell2021}, which are widely used to represent potential energy surfaces and their couplings. The parameters of a VC Hamiltonian can be obtained beforehand using electronic-structure theory. While here we address the challenge of simulating the dynamics generated by VC Hamiltonians, the parametrisation of the Hamiltonian can pose a separate challenge itself, which we discuss further below. Alternatively, electronic degrees of freedom can be included in the simulation using a fermion-to-qubit encoding without the need to parametrise a VC Hamiltonian~\cite{Ha2024}.
 
\begin{figure}[t!]
    \centering
    \includegraphics[]{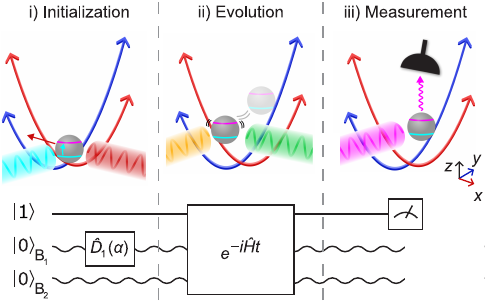}
    \caption{\textbf{Experimental implementation of an MQB simulator for chemical dynamics.} Three stages of MQB dynamics simulation: i) initialization, ii) evolution, and iii) measurement.
    Top row: trapped ion during each stage, with colors and symbols as in Fig.~\ref{fig:generaloverview}.
    Bottom row: circuit diagram of the pulse sequence acting on the qubit and two modes $B_1$ and $B_2$ (wavy lines). 
    \textbf{i)} During initialization, the qubit is prepared in state $\ket{1}$ and the vibrational modes are prepared in their ground states. Then, $B_1$ is displaced by $\hat{D}_1(\alpha)$ using a state-dependent force enacted by laser beams (cyan and red). 
    \textbf{ii)} During evolution, the Hamiltonian is applied for duration $t$ using state-dependent forces enacted by laser beams (yellow and green). 
    \textbf{iii)} The electronic state is measured using a photon counter through state-dependent fluorescence (pink).  
    } 
    \label{fig:MQBimplementation}
\end{figure}

Our simulation of molecular dynamics comprises three stages (Fig.~\ref{fig:MQBimplementation}). First, the initial wavefunction is prepared by exciting the qudit and displacing the relevant motional modes. Second, using laser-ion interactions with frequencies and intensities chosen to reproduce the molecular VC Hamiltonian, the simulator is evolved for some duration. Third, desired observables are measured. This process is repeated for varying evolution durations, allowing the reconstruction of observables as a function of time.

\begin{figure*}[t!]
    \centering
    \includegraphics[width=1.0\textwidth]{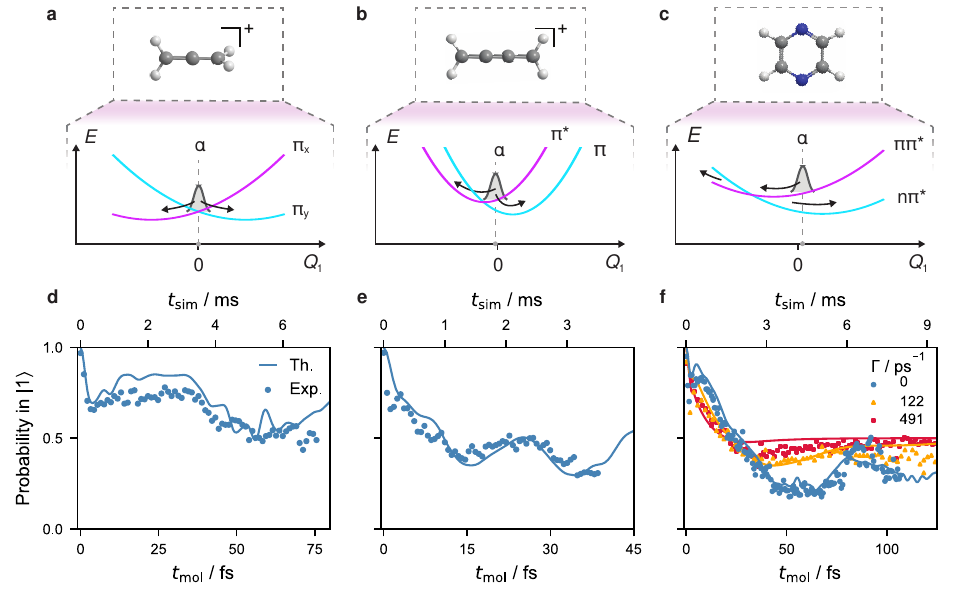}
    \caption{\textbf{Quantum simulation of photochemical dynamics in (a,d) photoionised allene, (b,e) photoionised butatriene, and (c,f) photoexcited pyrazine.} 
    \textbf{(a--c)}~Potential energy surfaces of each molecule (one-dimensional slice at $Q_2=0$; diabatic states in pink and cyan), with the wavepacket (gaussian width not to scale) initialised by displacement to $Q_1 = \alpha$. Each molecule results in a distinct energy landscape, where the conical intersection is (a) symmetric, (b) peaked, and (c) sloped. Black arrows indicate possible pathways for wavepackets to evolve on the PESs. 
    \textbf{(d--f)}~Corresponding molecules are simulated on an ion-trap MQB simulator using the LVC Hamiltonian of~\cref{eq:H_ion} with parameters from \cref{tab:molparam}. The probability of finding the wavepacket in the initial diabatic state as a function of the molecular evolution time (bottom axes).  
    Additionally, in (f), open-system dynamics of pyrazine under heating and cooling dissipation with varying rates $\Gamma$. Solid lines are the solutions of the Schr\"odinger equation for closed-system dynamics and of the master equation (\cref{eq:Lindblad}) for open-system dynamics. The simulator times shown in the top axes correspond to the closed-system dynamics. The scaling factors for the open-system dynamics with $\Gamma = \{122, 491\}\; \mathrm{ps}^{-1}$ are $F = \{1.69, 1.24\}\times 10^{-11}$.  Experimental data points are obtained by averaging over $M=500$ measurement results. Error bars of probabilities ($P$) are due to quantum projection noise, calculated from the binomial distribution as $\sigma = \sqrt{P(1-P)/M}$, and are smaller than the plot markers. 
    } 
    \label{fig:dynamics_figure}
\end{figure*}

We simulate the photoinitiated non-adiabatic dynamics in three molecules: photoionised allene (\ce{C3H4+}), photoionised butatriene (\ce{C4H4+}), and photoexcited pyrazine (\ce{C4N2H4}). The three molecules exhibit a wide variety of photochemical dynamics due to differences in their potential energy surfaces (see Fig.~\ref{fig:dynamics_figure}a-c). Allene is photoionised to a degenerate pair of $\pi$ states that are coupled via symmetry-breaking vibrational modes: bond alternation and torsion~\cite{woywod92}. The potential is thus symmetric about the conical intersection along both vibrational modes. Butatriene is photoexcited to the cationic $\pi$* state, coupled to the $\pi$ state through central bond stretching and torsion~\cite{cattarius01,ryabinkin14}. Both electronic potential-energy surfaces are displaced along the bond-stretching mode, leading to a peaked conical intersection, i.e., one where the lower adiabatic surface decreases in energy in all directions away from the intersection. Finally, pyrazine is photoexcited to the bright (large transition dipole moment) $\pi\pi$* state and decays to the dark $n\pi$* state, and its dynamics is dominated by a ring breathing mode and an out-of-plane hydrogen wag~\cite{woywod94,kuhl02}. The resulting Hamiltonian leads to a sloped conical intersection, where the lower adiabatic potential energy increases in one direction (here, $-Q_1$).

Each of the three molecules is well described by a linear VC (LVC) model, where the two electronic states and the two vibrational modes are linearly coupled. Denoting the initially excited electronic state as $\ket{1}$ (and the other one as $\ket{0}$), the Hamiltonian is~\cite{MacDonell2021}
\begin{multline}
    \hat{H}_\mathrm{mol} =  - \tfrac{1}{2}\Delta E \hat{\sigma}_z + \sum_j \omega_j \hat{a}_j^\dag \hat{a}_j  + {} \\ \frac{\kappa}{\sqrt{2}} \hat{\sigma}_z (\hat{a}_1^\dag + \hat{a}_1) + \frac{\lambda}{\sqrt{2}} \hat{\sigma}_x (\hat{a}_2^\dag + \hat{a}_2),
    \label{eq:LVCHam2}
\end{multline}
where $\hat{\sigma}_{x,z}$ are the Pauli operators and $\hat{a}_j$ is the annihilation operator of the $j$th mode with frequency $\omega_j$. $\Delta E$ is the energy difference between the two electronic states, while $\kappa$ and $\lambda$ are the tuning and coupling parameters, respectively. The initial wavefunction is a coherent state displaced along the mode $j=1$ by $\hat{D}_1(\alpha) = \exp(\alpha\hat{a}_1^\dag - \alpha^*\hat{a}_1)$. The parameter values for each molecule are given in~\cref{tab:molparam}.

\begin{table}
    \begin{tabular}{lccc}
        \toprule
        & Allene~\cite{woywod92} & Butatriene~\cite{ryabinkin14} & Pyrazine~\cite{kuhl02} \\ \midrule
        $\ket{1}$ & $\pi_x$ & $\pi^*$ & $\pi\pi^*$ \\ 
        $\ket{0}$ & $\pi_y$ & $\pi$ & $n\pi^*$ \\
        $\omega_1/2\pi$ (THz) & 22.5 & 62.9 & 17.9 \\ 
        $\omega_2/2\pi$ (THz) & 57.3 & 22.0 & 28.5 \\ 
        $\Delta E/2\pi$ (THz) & 0 & 131.5 & 199 \\ 
        $\kappa/2\pi$ (THz) & 74.7 & 62.1 & -30.7 \\ 
        $\lambda/2\pi$ (THz) & 67.7 & 69.6 & 63.3 \\ 
        $\alpha$ & 0 & $-0.140$ & 0.210 \\ 
        $F/\num{e-11}$ & 1.08 & 1.10 & 1.33 \\
        \bottomrule
    \end{tabular}
    \caption{\textbf{Parameters of the LVC Hamiltonian $\hat{H}_\mathrm{mol}$,} through which the MQB simulator can be programmed to simulate different molecules. $\ket{0}$ and $\ket{1}$ are the two relevant electronic states of the molecule. $\omega_1$ and $\omega_2$ are the frequencies of the two vibrational modes. $\Delta E$ is the energy difference between the electronic states at the origin. $\kappa$ and $\lambda$ are the tuning and coupling parameters, respectively. $\alpha$ is the displacement of the initial wavepacket. Parameters programmed on the MQB simulator are obtained by scaling the molecular frequencies by $F$. All molecular parameters are literature values obtained from electronic-structure calculations~\cite{woywod92, ryabinkin14, kuhl02}.}
    \label{tab:molparam}
\end{table}

\subsection{Trapped-Ion Simulation}

Each term in the LVC Hamiltonian $\hat{H}_\mathrm{mol}$ can be implemented on a trapped-ion MQB simulator consisting of a qubit (qudit with $d = 2$) representing the electronic states of the molecule and two motional modes of the ion in its harmonic confining potential representing the vibrations (Fig.~\ref{fig:generaloverview}e)~\cite{MacDonell2021}. This correspondence between molecular and trapped-ion Hamiltonian terms underpins the straightforwardness of our simulation approach. The vibronic couplings are implemented by tunable, laser-driven spin-dependent forces (SDF) in the Lamb-Dicke regime~\cite{MacDonell2021}, which drive the motional modes depending on the state of the qubit. After interaction-frame transformations and rotating-wave approximations~\cite{Wineland1998, MacDonell2021}, the Hamiltonian for interactions involving mode $j$ is
\begin{multline}
    \hat{H}_{j,\phi_s}^\mathrm{SDF}(\delta, \Omega) = \frac{\Omega}{2} 
    \left(\hat{\sigma}_x\cos\phi_s + \hat{\sigma}_y\sin\phi_s\right)(\hat{a}_j^\dagger + \hat{a}_j )  \\
    + \delta \hat{a}_j^\dagger \hat{a}_j,
    \label{eq:H_sdf_main}
\end{multline}
where $\hat{\sigma}_{x,y,z}$ are the Pauli matrices, $\phi_s$ is a phase that determines the qubit operator, which can be adjusted by the laser phase, $\Omega$ is the Rabi frequency that is tuneable with the laser intensity, and $\delta$ is the adjustable detuning of the laser frequency from the motional mode. We use the notation $\hat{H}_{j,x}^\mathrm{SDF}$ and $\hat{H}_{j,y}^\mathrm{SDF}$ for interactions where $\phi_s = 0$ and $\phi_s=\pi/2$, respectively. An SDF interaction in the $\hat{\sigma}_z$ basis can be obtained with a qubit basis rotation, $\hat{H}_{j, z}^\mathrm{SDF} = \hat{R}_x(\pi/2) \hat{H}_{j, y}^\mathrm{SDF} \hat{R}_x(-\pi/2)$, where $\hat{R}_x(\theta)$ are laser-driven qubit rotations around the $x$-axis of the Bloch sphere. We can also implement a Hamiltonian that generates a qubit rotation around $\hat{\sigma}_z$ by sandwiching a laser-driven interaction $\hat{H}_y^\mathrm{Q}(\chi) = \chi\hat{\sigma}_y/2$ between two $\hat{R}_x$ rotations, i.e., $\hat{H}_z^\mathrm{Q} = \hat{R}_x(\pi/2) \hat{H}_{y}^\mathrm{Q} \hat{R}_x(-\pi/2)$, where the Rabi frequency $\chi$ is tuneable using the laser's power.

$\hat{H}_\mathrm{mol}$ is experimentally implemented by simultaneously driving two SDF interactions and a qubit rotation, resulting in the ion Hamiltonian $\hat{H}_\mathrm{ion} = F \hat{H}_\mathrm{mol}$, where the scaling factor $F$ scales molecular frequencies (THz) and timescales (ps) to the trapped ion's frequencies (MHz) and timescales (ms). Overall,
\begin{multline}
    \hat{H}_\mathrm{ion} = \hat{H}_{1, z}^\mathrm{SDF}(F \omega_1, \sqrt{2}F\kappa) + \hat{H}_{2, x}^\mathrm{SDF}(F \omega_2, \sqrt{2} F \lambda) \\ + \hat{H}_z^\mathrm{Q}(-F \Delta E).
    \label{eq:H_ion}
\end{multline}
All parameters of~\cref{eq:H_ion} ($\omega_1$, $\omega_2$, $\kappa$, $\lambda$, and $\Delta E$) are programmable and can be set to simulate any two-state, two-mode LVC molecule. To obtain faster ion-trap dynamics, $F$ is maximised within the constraint of the available laser power~\cite{MacDonell2021,olayaagudelo2024arXiv}.

\section{Results}

\subsection{Closed-System Experiment}

We perform our experiment with a single \ce{^{171}Yb^+} ion electromagnetically held in vacuum with a quadrupole ion trap~\cite{MacDonell2023,Valahu2023}. A combination of radio-frequency and static electric fields confine the ion and give rise to secular vibrational motions with frequencies $\{\omega_x, \omega_y, \omega_z\} = 2\pi \times \{1.33, 1.51, 0.5\}$~MHz. These vibrations are harmonic to an excellent approximation. We encode the two molecular electronic states into the two magnetically insensitive hyperfine levels of the ion's \ce{^2S_{1/2}} ground state, with labels $\ket{0} \equiv \ket{F=0, m_F = 0}$ and $\ket{1} \equiv \ket{F=1, m_F=0}$. The molecular vibrational modes, $B_1$ and $B_2$, are encoded in the vibrational modes along the radial $x$ and $y$ directions.  

The experimental sequence for enabling the direct simulation of photochemical dynamics in the time domain is shown in Fig.~\ref{fig:MQBimplementation} (methods are detailed in~\cite{MacDonell2023, Valahu2023}). We program the experiment with parameters of $\hat{H}_\mathrm{ion}$ chosen according to~\cref{tab:molparam} to implement each of the three target molecules. In each case, the scaling factor $F$ is chosen to be $F = \Omega_1 / \sqrt{2}\kappa$, where $\Omega_1$ is the Rabi frequency of the SDF interaction of~\cref{eq:H_sdf_main} with mode $B_1$. The specific implementation of each simulation follows the three-stage process introduced above. First, initialization prepares the qubit in $\ket{1}$ and the modes in their ground states, followed by displacing $B_1$ by $\hat{D}_1(\alpha)$. Second, the system is evolved by applying $\hat{H}_\mathrm{ion}$ for an experimentally variable duration $t_\mathrm{ion}$, which is related to the molecular timescale by $t_\mathrm{mol} = F t_\mathrm{ion}$. Third, measurement of the electronic populations is achieved by making a $\hat{\sigma}_z$ measurement on the qubit through state-dependent fluorescence~\cite{Olmschenk2007}. 

Experimental measurements of the probability of finding the wavepacket in the initial diabatic state as a function of evolution time are shown in Fig.~\ref{fig:dynamics_figure}d-f. For all three molecules, experiments match the predicted dynamics arising from conventional calculations. Small discrepancies are due to miscalibrated experimental parameters and hardware noise. In all cases, we observe a population decay due to the strong coupling between electronic states in the region of the initial wavefunction. We further observe distinct behaviors for each molecule, caused by the different potential-energy surfaces. The symmetric conical intersection of the allene cation causes a rapid decay of the population because the initial state is at the point of strongest coupling between electronic states; however, only a small amount of population is transferred due to the degeneracy of the $\pi$ states. The initial geometry of allene is located at a region of strong coupling between electronic states, which causes a rapid initial decay; however, the $\pi$ states are energetically equivalent, so no more than half the population is transferred. The butatriene cation likewise has an initial state with strong coupling, but the lower potential energy minimum of the $\pi$ state favours a greater overall population transfer. Finally, the location of the pyrazine $\pi\pi$* minimum near its intersection with the n$\pi$* state leads to a near-complete occupation of the n$\pi$* state; however, the greater distance to the intersection leads to slower initial dynamics.

\subsection{Open-System Experiment}

To further demonstrate the versatility of the MQB approach, we perform a quantum simulation of open-system vibronic dynamics, which enables the simulation of environmental molecular conditions.
To do so, we externally inject noise into the same single-ion simulation above. Doing so shows that what is classically a harder computational problem can be solved using the same quantum resources on an MQB simulator~\cite{olayaagudelo2024arXiv, Kim2022}.

In particular, we simulate the non-equilibrium ultrafast dynamics of a pyrazine molecule interacting with a thermal bath. The dissipation is simulated by simultaneously injecting heating and cooling of the vibrational modes, described by the Lindblad master equation~\cite{Henkel1999, Brownnutt2015},
\begin{equation}
   \frac{d\hat{\rho}}{dt} = -i[\hat{H}_\mathrm{ion}, \hat{\rho}] + \sum_j \left( \gamma_{+,j} \mathcal{D} [ \hat{a}^\dag_j ] \hat{\rho}+ \gamma_{-,j} \mathcal{D} [\hat{a}_j] \hat{\rho}\right),
   \label{eq:Lindblad}
\end{equation}
where $\mathcal{D}[\hat{a}^\dag_j]$ is the heating dissipator for mode $j$,
\begin{equation}
    \mathcal{D}[\hat{a}^\dag_j]\hat{\rho} = \hat{a}^\dag_j {\hat{\rho}} \hat{a} - \tfrac{1}{2}\{\hat{a}^\dag_j\hat{a},{\hat{\rho}}\},
    \label{eq:LindbladRates}
\end{equation}
and similarly for the cooling dissipator $\mathcal{D}[\hat{a}]\hat{\rho}$. Heating and cooling have rates $\gamma_{+,j}$ and $\gamma_{-,j}$, respectively, and describe energy transfer from the environment to the vibrations (heating) or vice versa (cooling). 

The dissipators are engineered in the simulator by injecting a noisy electric field at the ion's location~\cite{Valahu2024arxiv}. The electric field noise results in cooling and heating dissipators with approximately equal rates, $\gamma_j = \gamma_{+,j} = \gamma_{-, j}$, corresponding to a thermal bath with infinite temperature~\cite{Henkel1999, Brownnutt2015}. In the molecule, this corresponds to a scaled dissipation rate $\Gamma_j = F^{-1} \gamma_j$. A noisy voltage signal, generated using an arbitrary waveform generator, is capacitively coupled onto a radial compensation electrode located $\SI{4.8}{mm}$ from the ion's position. The signal contains two frequency components oscillating near the vibrational mode frequencies, $\omega_x$ and $\omega_y$. The rates $\gamma_1$ and $\gamma_2$ are programmed by varying the corresponding signal amplitudes. The decay rates are calibrated using the trapped ion through a standard sideband thermometry experiment~\cite{Monroe1995, Turchette2000}.

The experiment correctly simulates the open-system dynamics at varying dissipation rates (Fig.~\ref{fig:dynamics_figure}f), showing the suitability of MQB simulation for the particularly challenging task of modelling ultrafast dynamics in open molecular systems. As expected, stronger dissipation leads to a faster loss of coherent features. With dissipation, the electronic population decays at long times to 0.5, consistent with a Boltzmann distribution at high temperature, where the populations of the two electronic states obey $n_1/n_0 = \exp(-\Delta E/k_\mathrm{B}T) = 1$. This equilibration occurs faster with stronger dissipation.

\section{Discussion}

In summary, we performed quantum simulations of non-adiabatic chemical dynamics of three molecules. We simulated the evolution of their vibrational wavepackets through conical intersections, obtaining distinct dynamics linked to the characteristics of their potential energy surfaces. We also simulated open-system dynamics of pyrazine coupled to a thermal bath using the same trapped-ion system.

Our experiment demonstrates three advantages of MQB simulations: programmability, resource efficiency, and the ability to simulate open systems.

MQB simulators are programmable, allowing us to simulate the dynamics of very different molecules simply by adjusting experimental parameters to match the different VC Hamiltonian in the same experimental apparatus.This programming is carried out through software-updated experimental parameters, without the need for hardware modification. Different features of photochemical processes were captured by the same MQB simulator: dynamics between different types of electronic states (ground or excited, singlet or doublet, degenerate or non-degenerate), vibrational modes and vibronic couplings with different symmetries, conical intersections with different topography (symmetric, peaked, and sloped), and a wide range of vibrational frequencies, vibronic couplings and wave-packet displacements.

MQB simulators require orders of magnitude lower quantum-hardware resourcing compared to qubit-only quantum simulations~\cite{MacDonell2021,Kassal2008}. We used a single trapped ion and a single laser pulse to simulate the ultrafast molecular dynamics of LVC models with two electronic states and two vibrational modes. The comparable quantum memory requirements can be estimated from the dimension of the necessary Hilbert space. Our simulations can be reproduced using 32 Fock states per mode, meaning that a comparable qubit-based simulation would need 11 qubits: $\lceil \log_2 32 \rceil = 5$ qubits for each of the two modes and an additional qubit for the electronic states. The CNOT gate requirements can be estimated using the resource-efficient Gray-code qudit-to-qubit encoding~\cite{Sawaya2020} with accuracy chosen to reproduce the mean-squared error of our MQB simulation (which is 0.0034, averaged over the duration of our simulation). Doing so with a first-order Suzuki-Trotter decomposition needs 1000 errorless CNOT gates for each of 300 Trotter steps. Therefore, a comparable qubit-based simulation could be achieved with $3 \times 10^5$ CNOT gates. A realistic quantum computer with noisy gates or quantum error correction would require even more qubits and gates. A resource estimation for an MQB simulation of lattice gauge fields found a similar reduction of over five orders of magnitude of quantum resources~\cite{crane2024}.

We also demonstrate the ability of MQB simulators to simulate open-system dynamics by injecting controllable noise without using additional quantum resources. Comparable qubit-only simulations would have required additional ancilla resources to simulate the environment. On the MQB simulator, by contrast, the classically more difficult open-system problem is an easier task because some of the native noise can be used in the simulation, allowing for longer simulations with greater accuracy~\cite{olayaagudelo2024arXiv}.

Extending MQB simulation beyond proof-of-principle experiments will enable increasingly challenging dynamics simulations. We envision opportunities for improvement in three areas: scale, non-linearities, and open quantum systems.

Scaling the experimental system from $1$ to $N$ trapped ions could enable the control of $3N$ motional modes~\cite{MacDonell2021}. The chief challenge to doing so is retaining the high quality of motional control over the additional modes, a matter complicated by spectral crowding and reduced interaction strengths~\cite{Wineland1998}. These challenges could be mitigated using the same  strategies developed for trapped-ion quantum computers using long ion chains, including advanced quantum-control methods and improved optical setups~\cite{Leung2018, Milne2020}.

MQB simulators could also be extended to implement higher-order vibronic-coupling models~\cite{MacDonell2021} or anharmonic potential energy surfaces. With the addition of beam-splitter interactions, the coherent laser-driven interactions demonstrated here are sufficient for universal control of MQB systems~\cite{Sutherland2021,Liu2024}, meaning that they can be composed into pulse sequences using quantum-control methods to engineer non-linear couplings or anharmonic potentials. However, full scope of this engineering remains an open question, because a fully general decomposition may scale unfavourably for high-dimensional, strongly coupled anharmonic potentials.  

Finally, MQB simulations could be extended to include other types of dissipation for more comprehensive and fully programmable open-system quantum dynamics~\cite{olayaagudelo2024arXiv}. The most significant sources of noise in quantum-simulation experiments could be used as a resource to simulate the common forms of molecular decoherence and dissipation, thus significantly extending the useful lifetime of the simulation~\cite{olayaagudelo2024arXiv}. Fully exploiting all available sources of noise would be a powerful simulation technique because of the difficulty of open-system chemical-dynamics simulations on classical computers. 

Combining these three types of improvements in future experiments would allow addressing one of the great challenges of computational chemistry: simulation of the ultrafast dynamics of large, complicated molecules in the condensed phase. In doing so, MQB simulators could outperform other quantum simulation approaches: we estimate that an MQB simulator with 20--30 trapped ions could perform quantum-chemical-dynamics simulations that are classically intractable. 

Our approach also opens up possibilities for hybrid simulation approaches that exploit complementary advantages of different types of hardware. For example, while an MQB simulator could carry out the dynamics, the electronic-structure calculations necessary to parametrise the VC Hamiltonians could be carried out on a classical computer or a digital quantum computer equipped with electronic-structure software.

\section*{acknowledgments}
We were supported by the U.S. Office of Naval Research Global (N62909-24-1-2083), the U.S. Army Research Office Laboratory for Physical Sciences (W911NF-21-1-0003), the U.S. Air Force Office of Scientific Research (FA2386-23-1-4062), the Wellcome Leap Quantum for Bio program, the Australian Research Council (FT220100359, FT230100653), Lockheed Martin, the Sydney Quantum Academy (VCOA, ADR, MJM), the University of Sydney Postgraduate Award scholarship (VGM), the Australian Government Research Training Program (FS), the Sydney Horizon Fellowship (TRT), and H.\ and A.\ Harley.

\section*{Data availability}

Experimental data in Fig.~3 is available at \url{https://doi.org/10.5281/zenodo.13948671}~\cite{navickas2024}.

% \bibliography{bib}


\end{document}